\documentclass[conference]{IEEEtran}

\IEEEoverridecommandlockouts
\usepackage{cite}
\usepackage{amsmath,amssymb,amsfonts}
\usepackage{algorithmic}
\usepackage{textcomp}
\usepackage{xcolor}

\usepackage{ascmac}
\usepackage[dvipdfmx]{graphicx}
\usepackage{multirow}
\usepackage{booktabs}
\usepackage{url}
\usepackage[normalem]{ulem}
\usepackage{color}
\useunder{\uline}{\ul}{}

\def\BibTeX{{\rm B\kern-.05em{\sc i\kern-.025em b}\kern-.08em
T \ kern-.1667em \ lower.7ex \ hbox {E} \ kern-.125emX}}
\begin{document}

\newcommand{\argmin}{\mathop{\rm arg~min}\limits}
\newcommand{\argmax}{\mathop{\rm arg~min}\limits}
\newcommand{\kmax}{\mathop{\rm max}\limits}

\newcommand{\termIndex}{\ensuremath{t}}
\newcommand{\docIndex}{\ensuremath{d}}
\newcommand{\nDocs}{\ensuremath{N}}
\newcommand{\nWords}{\ensuremath{n_\term}}
\newcommand{\tfidf}{\ensuremath{tf\mbox{-}idf (\termIndex , \docIndex)}}
\newcommand{\tf}{\ensuremath{tf (\termIndex, \docIndex)}}
\newcommand{\idf}{\ensuremath{idf (\termIndex)}}
\newcommand{\df}{\ensuremath{df(\term , \doc)}}

\newcommand{\training}{\ensuremath{L}}
\newcommand{\testing}{\ensuremath{V}}
\newcommand{\dfTraining}{\ensuremath{df_\training (\termIndex)}}
\newcommand{\nDocsTraining}{\ensuremath{\nDocs_\training}}

\newcommand{\elmt}{\ensuremath{x}}
\newcommand{\vect}{\ensuremath{{\bf x}}}
\newcommand{\ltvec}{\ensuremath{\vect_{L2}}}

\newcommand{\rowi}{\ensuremath{i}}
\newcommand{\coli}{\ensuremath{j}}
\newcommand{\elij}{\ensuremath{x^{(\rowi , \coli)}}}
\newcommand{\elijnorm}{\ensuremath{\elij_{norm}}}
\newcommand{\xnorm}{\ensuremath{\elmt_{norm}}}
\newcommand{\xmin}{\ensuremath{\elmt_{min}(\coli)}}
\newcommand{\xmax}{\ensuremath{\elmt_{max}(\coli)}}

\newcommand{\weight}{\ensuremath{w}}
\newcommand{\wi}{\ensuremath{\weight^{(\coli)}}}

\newcommand{\nclass}{\ensuremath{n}}
\newcommand{\class}{\ensuremath{c}}
\newcommand{\nFeatures}{\ensuremath{m}}

\newcommand{\xvar}{\ensuremath{X_{var}}}

\newcommand{\ftuple}{\textit{5-tuple}}

\newcommand{\x}{\ensuremath{x}}
\newcommand{\X}{\ensuremath{X}}
\newcommand{\xSpace}{\ensuremath{{\cal X}}}
\newcommand{\y}{\ensuremath{y}}
\newcommand{\ySpace}{\ensuremath{{\cal Y}}}
\newcommand{\Y}{\ensuremath{Y}}
\newcommand{\nClasses}{\ensuremath{K}}
\newcommand{\predY}{\ensuremath{\hat{\y}}}
\newcommand{\threshold}{\ensuremath{\tau}}
\newcommand{\rejectSymbol}{\ensuremath{\phi}}
\newcommand{\score}{\ensuremath{S}}
\newcommand{\scoreY}{\ensuremath{\score_{\y}}}
\title{
IDPS Signature Classification with a Reject Option \\ and the Incorporation of Expert Knowledge
}
\author{\IEEEauthorblockN{Hidetoshi Kawaguchi$^{1,2}$, Yuichi Nakatani$^1$ and  Shogo Okada$^{2}$}
\IEEEauthorblockA{$^1$\textit{Network Innovation Center, Nippon Telegraph and Telephone(NTT), Tokyo, Japan} \\
\textit{$^2$School of Information Science, Japan Advanced Institute of Science and Technology (JAIST), Ishikawa, Japan}\\
\{hidetoshi.kawaguchi.my, yuichi.nakatani.rd\}@hco.ntt.co.jp, \{kawa.hide39, okada-s\}@jaist.ac.jp}
}

\maketitle

 \begin{abstract}
  As the importance of intrusion detection and prevention systems (IDPSs) increases, great costs are incurred to manually manage the signatures that are generated by malicious communication pattern files.
  Experts in network security management need to classify signatures by importance for an IDPS to work optimally.
  In this study, we propose and evaluate a machine learning signature classification model with a reject option to reduce the cost of setting up an IDPS.
  To train the proposed model, it is essential to design features that are effective for signature classification.
  Experts classify signatures with predefined if-then rules.
  An if-then rule returns a label of low, medium, high, or unknown importance based on keyword matching of the elements in the signature.
  Therefore, we first design two types of features, symbolic features (SFs) and keyword features (KFs), which are used in keyword matching for the if-then rules.
  Next, we design web information and message features (WMFs) to capture the properties of signatures that do not match the if-then rules.
  The WMFs are extracted as term frequency-inverse document frequency (TF-IDF) features of the message text in the signatures.
  The features are obtained by web scraping from the referenced external attack identification systems described in the signature.
  Because failure needs to be minimized in the classification of IDPS signatures, as in the medical field, we consider introducing a reject option in our proposed model.
  The effectiveness of the proposed classification model is evaluated in experiments with two real datasets composed of signatures labeled by experts:
  (i) a dataset that can be classified with if-then rules and (ii) a dataset with elements that do not match an if-then rule.
  In the experiment, the proposed model is evaluated from two perspectives: classification accuracy and reject option performance.
  In both cases, the combined SFs and WMFs performed better than the combined SFs and KFs.
  We also show that using an ensemble of neural networks improves the performance of the reject option.
  An analysis shows that experts refer to natural-language elements in the signatures and information from external information systems on the web.
 \end{abstract}


\section{Introduction}
\footnote{This work has been submitted to the IEEE for possible publication. Copyright may be transferred without notice, after which this version may no longer be accessible.}
Intrusion detection and prevention systems (IDPSs) are security systems for computer networks.
These systems perform several actions, for example, recording communication logs, alerting to the need for investigation, and preventing attacks on a network, when they detect malicious communications.
The IDPS detects intrusions based on signatures, which are malicious communication pattern files \cite{Sdas}.
We can set the actions of the IDPS for each signature.

Experts in network security management need to determine whether each action should be executed for signatures based on the importance labels assigned to each signature.
Great costs are incurred to manage signature data manually.

To reduce the cost of classifying signatures,
this study aims to build a signature classification model with a reject option (RO) using machine learning.
The RO is a function used to cancel classification when the model's output appears uncertain.
Signature classification can cause catastrophic damage to communication networks if it fails.
We should consider the cost of misclassification to be as high as in the medical field.
It is a natural idea to add an RO to prevent classification errors as well as possible.
To design features for classification, we refer to 
methods by which experts classify importance.
In general, experts classify signatures somewhat automatically.
First, the experts classify signatures using an if-then rule that they construct.
The if-then rule returns an importance label or an unknown label according to the results of keyword matching on the elements in a signature.
The experts then manually classify signatures that are determined to be unknown by the if-then rule.

In this study, a single classification model is used to classify all signatures.
Therefore, following the expert classification methods described above, we propose three sets of features.
The first is a feature set that is subject to the conditions of an if-then rule.
The second set is a feature set obtained based on the keywords in the message text and the attack identification system name.
The third set is a language feature extended by web scraping from the messages and external attack identification systems described in the signatures.

We evaluate the effectiveness of the proposed model using two datasets, which consist of real signatures labeled by experts.
One is a dataset that can be classified with an if-then rule.
The other dataset consists of signatures that do not match the if-then rule.
The experiments show that the proposed features are valid for the task of classifying the two datasets by using multiple machine learning classification models.

The contributions of this paper are summarized as follows:
(i) We develop a real dataset for signature classification with experts in network security operational practice.
(ii) We design an effective feature set to classify the signature patterns in classification tasks with an RO.
(iii) We evaluate both the classification accuracy and the RO of the trained models.
(iv) The analysis identifies and experimentally demonstrates practical machine learning algorithms for ROs and signature elements of interest to experts.

\section{Related works}
\subsection{IDPS}
Much research has been performed on signature management.
Stakhanova et al. proposed an analytical model for identifying inconsistent signatures \cite{Stakhanova:2010:MID:1752046.1752051}.
They represented a signature as a nondeterministic finite automaton and identified overlapping signatures based on automaton equivalence.
For the same purpose, Massicotte and Labiche proposed a different approach that utilizes set theory and automata theory \cite{F5958211}.

Shahriar and Bond proposed a method for automatically generating signatures \cite{H8328450}.
The genetic algorithm generates them based on other signatures.
For the same purpose, Lee et al. applied latent Dirichlet allocation, and Fallahi et al. applied decision trees \cite{S7569096,N7585840}.

There are several studies on classification models of malignant communication that use machine learning.
The input to these classification models is the set of features that express the characteristics of a communication.
The output is the category of malignant communication.
Methods such as support-vector machines (SVMs) and  decision treeshave been applied to the classification of malignant communication \cite{Constantinides, Chandre}.

Cort\'{e}s and G\'{o}mez analyzed two approaches that considered false-alarm minimization and alarm correlation techniques and reduced false alarms \cite{F8809121}.
Tadeusz also proposed a system that uses machine learning to minimize false alarms \cite{P10.1007/978-3-540-30143-1_6}.
Although it is different from the approach for maintaining signatures, the research purpose, which is to reduce the burden of security operations, is the same as that of our research.

To the best of our knowledge, the idea of performing automatic classification by using machine learning to classify signatures directly has not been explored.
In this paper, the primary challenge in classifying signatures in machine learning is addressed.
We design effective features for classification and evaluate and analyze them.

\subsection{Reject option (RO)} \label{subsec:reject-option}
The RO was pioneered by Chow \cite{Chow1957,Chow1970}.
The idea of dropping a classification according to certain criteria is widely used, and these studies are reviewed.

Hanczar et al. proposed a method combining SVM and one-class SVM to improve the RO performance.
Harish et al. performed a theoretical analysis of the RO in the case of three or more classifications \cite{Harish2018}.
Goepfert et al. extended the self-adjusting memory architecture (SAN-KNN) \cite{Losing2016} and adaptive random forest (ARF) \cite{Gomes2017} methods to incorporate an RO for classification in nonstationary environments \cite{Geoepfert2018}.

The RO is practical and has been studied for certain applications, especially in the medical field.
Waseem et al. constructed and analyzed a classification model with an RO that predicts cancer based on genetic information \cite{Waseem2019}.
Lin et al. proposed a classification model with an RO to classify biomedical images \cite{Lin2018}.
Their model uses SVM, which is an ingenious way to calculate confidence.
The confidence is based on the average ratio of the distance to the separating hyperplane during classification in the SVM, normalized to $[0,1]$, and the distance to the centroid of each class in the feature space.
Raghu et al. showed theoretically and experimentally that for images that are automatically classified by a classification model, the uncertainty of medical images can be directly estimated to determine whether to seek a second opinion \cite{Raghu2019}.
An RO is helpful in fields where the impact of misclassification is significant, such as the medical field.

In IDPS signature classification, the subject of this study, the cost of failure is as high as in medical fields.
This is because malicious communications may be missed due to misconfiguration of the IDPS and lead to security incidents.
Security incidents should be avoided because they damage public trust in the organization operating the network.
If regular communication is interrupted, the convenience of the network is also reduced.
To mitigate such risks, using an RO in the classification model of IDPS signatures is a natural solution.

\section{Problem setup}
\subsection{Dataset}

\begin{table}[t] \centering
  \caption{dataset of signatures}
  \label{table:dataset}
\begin{tabular}{|c||c|c|c|c|}
\hline
\multirow{2}{*}{datasets}& \multicolumn{3}{c|}{importance levels}                                                & \multirow{2}{*}{sum}         \\ \cline{2-4}
                         & low                         & \multicolumn{1}{c|}{medium} & \multicolumn{1}{c|}{high} &                              \\ \hline \hline
AAD                      & \multicolumn{1}{r|}{$3,936$} & \multicolumn{1}{r|}{$93$}  & \multicolumn{1}{r|}{$436$}& \multicolumn{1}{r|}{$4,465$} \\ \hline
MAD                      & \multicolumn{1}{r|}{$1,119$} & \multicolumn{1}{r|}{$122$} & \multicolumn{1}{r|}{$59$} & \multicolumn{1}{r|}{$1,300$} \\ \hline
\end{tabular}
\end{table}

The evaluations in this paper are carried out on signatures labeled by experts engaged in network security management.
The experts design if-then rules to classify as many signatures as possible.
An if-then rule returns a label of low, medium, high, or unknown importance based on keyword matching of the elements in the signature. 
First, the experts automatically classify signatures with the if-then rules.
Next, the experts manually determine the labels of signatures that do not match the if-then rules.
Two datasets are then created: one for signatures classified by the if-then rule and the other for signatures classified manually.
The former is called the automatically annotated dataset (AAD), and the latter is called the manually annotated dataset (MAD).

Table \ref{table:dataset} shows the number of AAD and MAD samples prepared by the experts.
Each signature is assigned one of three importance labels: low, medium, or high.
Based on the importance level, the experts set an action of the IDPS for communications that match the signature.

\subsection{Notation of signatures}

\begin{figure}[t] \centering
 \includegraphics[width=9cm]{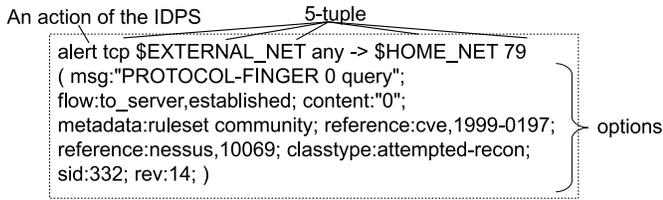}
 \vspace{-0.5cm}
 \caption{A specific example of IDPS signatures}
 \label{fig:signature}
\end{figure}

The signatures in the AAD and MAD are written in the notation of the IDPS engine called Snort \footnote{\url{https://snort.org/}}.
Fig. \ref{fig:signature} shows a concrete example\footnote{\url {https://www.snort.org/downloads/#rule-downloads}}.
The first word ``alert'' is the action taken by the IDPS when the signature is matched.
Because the experts set up actions based on importance, actions cannot be entered into the classification model.
Features are extracted from the strings after the action.

``tcp \$EXTERNAL\_NET any -$>$ \$HOME\_NET 79'' is a 5-tuple of values.
The 5-tuple is a set of five values listed in the header of an IP packet.
``tcp'' is the communication protocol.
``\$EXTERNAL\_NET'' is the source IP address.
``any'' is the source port number.
``\$HOME\_NET'' is the destination IP address.
``79'' is the destination IP address.

A 5-tuple is an essential element of signatures.
A string in parentheses after 
the 5-tuple 
is optional.
The options are expressed in key-value format with the following conditions:
The key and value are linked with colons.
Depending on the key, there may be more than one value.
Some values do not have a key, such as \textit{nocase}.

We focus on four elements in the options: \textit{msg} (abbreviation for message), \textit{metadata}, \textit{reference}, and \textit{classtype}.
The if-then rule classifies signatures in terms of 5-tuples and these elements alone.

\textit{msg} is a string written to a log or alert when a signature is matched with a communication.
``PROTOCOL-FINGER 0 query'' in Fig. \ref{fig:signature} is an example of this.

\textit{metadata} is an element that represents information in the key-value format.
The ``ruleset community'' in Fig. \ref{fig:signature} is an example of metadata.
A space separates the key and value.
Commas delimit key-value sets.
This example has one key-value set.

\textit{reference} describes a pointer to external information about the attack identification system.
In Fig. \ref{fig:signature}, it is described as ``cve,1999-0197''.
In this example, \textit{reference} refers to Common Vulnerabilities and Exposures (CVE) 1999-0197.

\textit{classtype} is a general group of malicious communications indicated by signatures.
In Fig. \ref{fig:signature}, it is ``attempted-recon''.
The groups that \textit{classtype} indicates are different from the importance levels judged by the experts.

\subsection{If-then rule}
The if-then rule classifies signatures by 
matching keywords and combinations of keywords.
Keyword matching is used to determine whether a word is included in a signature.
The key-value pairs used in keyword matching are the 5-tuple, \textit{msg}, \textit{metadata}, \textit{reference}, and \textit{classtype}.
Keyword matching for \textit{metadata} uses a key-value pair as one keyword.
\textit{msg} keyword matching does not consider word position.
In other words, keyword matching for \textit{msg} determines whether a certain word appears.
Keyword matching for \textit{reference} determines whether a specific system is referred to, and it does not use an ID.
The elements of the 5-tuple are extracted and judged individually.

The if-then rule assigns importance labels only to signatures that match certain conditions.
Signatures that do not qualify for keyword matching are assigned the label ``unknown''.
\section{A classification model with a reject option}
 \begin{figure*}[t] \centering
 \includegraphics[width=16cm]{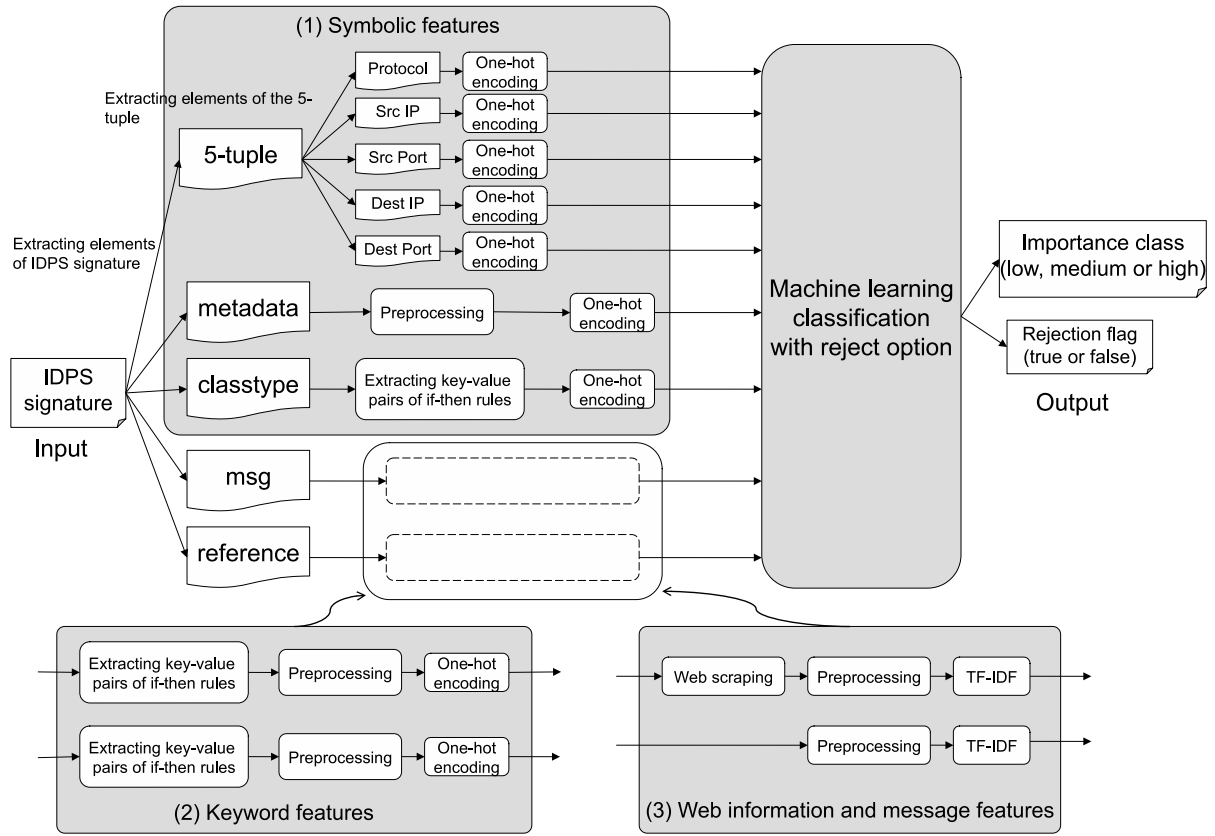}
  \caption{Classification model with an RO for IDPS signatures. There are three types of features: (1) \textit{symbolic features}, (2) \textit{keyword features} and (3) \textit{web information and message features}.}
 \label{fig:feature-engineering}
  \end{figure*}
To construct a classification model with an RO, we propose \textit{symbolic features} (SFs), \textit{keyword features} (KFs), and \textit{web information and message features} (WMFs).
We also describe the behavior of the RO incorporated into the classification model.
Fig. \ref{fig:feature-engineering} shows the procedure of the proposed classification model.
The SFs and KFs are designed with reference to if-then rules.
The WMFs are designed with reference to interviews with experts.

\subsection{Symbolic features}
SFs are extracted from the \textit{5-tuple}, \textit{metadata}, and \textit{classtype}, each of which is extracted as a feature with one-hot encoding.
The top of Fig. \ref{fig:feature-engineering} shows this procedure.

The \textit{classtype} is directly extracted as a feature using one-hot encoding.
However, the 5-tuple and \textit{metadata} need to be preprocessed.
The 5-tuple is separated into its five values.
After that, each value is converted into features by one-hot encoding.
In the extraction procedure for \textit{metadata}, all the key-value pairs are extracted first.
Then, all the extracted key-value pairs are reordered and combined into a string to form a single symbol.

\subsection{Keyword features}
KFs are designed for keyword matching on \textit{msg} and \textit{reference} in the if-then rule.
The lower left part of Fig. \ref{fig:feature-engineering} shows a procedure for extracting KFs.
KFs are extracted from \textit{msg} and \textit{reference} according to the presence or absence of the keywords in the if-then rule.
After extraction, they are converted with one-hot encoding.

To convert \textit{msg} to features, a list is created from the words used in the matching conditions from the if-then rule.
Sets of words in the list are extracted as symbols from \textit{msg} in the order of the list.
If no word matches the word list, a dummy symbol is used to indicate that the word does not exist.
Then, the symbols are converted with one-hot encoding.

To extract features from \textit{reference}, a list is made from system names that exist in the if-then rule.
The system names pointed to by \textit{reference} are combined and treated as a symbol.
If a system name does not match the list, it is extracted as a dummy symbol indicating this.
These symbols are converted to features by one-hot encoding.

\subsection{Web information and message features }
Many signatures cannot be classified by an if-then rule.
We need to add a new criterion to capture the properties of such signatures.
We interviewed experts to design new features from the criteria used by experts to classify features manually.
From the interviews, we learned the following:
\begin{itemize}
\item The experts check all the information in \textit{msg}.
\item The experts check the external information of \textit{reference} via a web search.
\end{itemize}
We assume that the whole \textit{msg} and the information on the web indicated by \textit{reference} are essential and propose features that can effectively use them.
To expand on this information, we apply term frequency-inverse document frequency (TF-IDF), which is frequently used in natural language analysis \cite{A8723825,Y8253040,P8250358}.
Web scraping is also used for feature extraction on \textit{reference}.
The lower right part of Fig. \ref{fig:feature-engineering} shows the procedure for converting to WMFs.

For \textit{reference}, web scraping is performed to obtain information from external references.
\textit{reference} is a set of names of vulnerability lists (CVE, Bugtraq, etc.) and their IDs or URLs, which allow information related to the signature to be uniquely identified.
For example, when referring to CVEs, information can be obtained by searching for an ID in web systems such as the National Vulnerability Database (NVD)\footnote{\url{https://nvd.nist.gov/vuln}} and RedHat's CVE Database\footnote{\url{https://access.redhat.com/security/security-updates/#/cve}}.
Examples of signature-related information include the software targeted by the malicious communication indicated by the signature and its version information.
The developer of the classification model needs to describe the web scraping process for each web system that publishes the information referred to by the reference.
Although applying the procedure to all web systems is difficult, it is possible to describe the procedure by focusing on 
frequently used web systems.
In what follows, \textit{reference} refers to information obtained by web scraping.

Before extracting TF-IDF values, \textit{msg} and \textit{reference} are cleaned as follows:
Since only alphabets, numbers, and underscores are used, other symbols are replaced by blanks.
Stop words \cite{nothman-etal-2018-stop} and words that appear only once in all signatures are removed.

The cleaned \textit{msg} and \textit{reference} are converted into feature vectors by TF-IDF separately.
Let $\docIndex$ be an identifier for a text (\textit{msg} or \textit{reference} in a signature) and $\termIndex$ be an identifier for a word; the TF-IDF is as follows:
\begin{equation}
 \tfidf = \tf \cdot \idf
\end{equation}
$\tf$ represents the number of occurrences (an integer greater than or equal to 0) of the word $\termIndex$ in the text $\docIndex$.
$\idf$ is calculated as follows:
\begin{equation}
 \idf = \log \frac{\nDocsTraining + 1}{\dfTraining + 1} + 1.
\end{equation}
$\nDocsTraining$ is the number of texts in the training data.
$\dfTraining$ is the number of texts in which the word $\termIndex$ appears among the $\nDocsTraining$ training samples.
In other words, the IDF used when converting the test data to feature vectors with TF-IDF is the IDF calculated in the training data.
When converting to TF-IDF, all words are treated as unigrams.
After conversion to TF-IDF, L2 normalization is performed for each WMF.
After L2 normalization, min-max scaling is performed with a minimum value of zero and a maximum value of one.

\subsection{If-then rule features and manual classification features}
We extract two types of feature sets.
One concatenates the SFs and KFs into a vector directly.
These connected features are called if-then rule features (ITRFs).
The ITRF is a feature design based on the if-then rule.
Second, we directly concatenate the SFs and WMFs to form a vector.
The connected features are called manual classification features (MCFs).
The MCF is a feature design considering manual classification by experts.

\subsection{Classification with a reject option}
The proposed model is equipped with an RO.
The RO means that the classification can be rejected depending on the value of the class's prediction score.

The RO can be used on any classification model as long as a prediction score can be calculated.
It is formulated as follows:
Let $\x \in \xSpace$ be the input class and $\y \in \ySpace = \{1,..,\nClasses\}$ be the output class.
Let $\scoreY : \xSpace \rightarrow \mathbb{R}$ be a function that computes the prediction score of class $\y$ for a classifier.
The classification model with an RO determines the final classification prediction, class $\predY$, as follows:
\begin{equation}
 \predY = \left\{
           \begin{array}{ll}
            \argmax_{\y \in \ySpace} \scoreY(\x)  & \mbox{if} \kmax_{\y \in \ySpace} \scoreY(\x) \geq \threshold \\
            \mbox{reject} & \mbox{otherwise.}
           \end{array}
          \right.
\end{equation}
$\threshold$ is the threshold, which is the hyperparameter of the RO.
\section{Experiment}
\subsection{Experimental setting}
\subsubsection{Outline of the experiment}
In this section, we confirm the classification accuracy and RO performance of the proposed classification model and analyze the validity of the feature design.
Specifically, experiments are conducted on the following process:
(1) We validate the proposed features (ITRFs and WMFs) on several traditional machine-learning models and compare the classification accuracy.
(2) We evaluate the quantified RO performance with accuracy-rejection curves (ARCs) and the area under the ARC (AU-ARC) \cite{Sajjad2009} and evaluate the classification accuracy.
(3) We explore the performance improvement due to the RO by using a robust machine-learning model for the RO (deep ensemble \cite{Lakshminarayanan2017}), which is said to better represent uncertainty.
(4) We analyze the importance of the proposed features to identify the signature elements that the experts regard as important for evaluating the signature.
The results of these experiments are described in the form given in Sections \ref{subsubsec:ca},\ref{subsubsec:ro},\ref{subsubsec:de},\ref{subsubsec:ae}.

\subsubsection{Evaluation method}
Due to the imbalanced dataset, we measure the balanced accuracy as the classification accuracy.
In addition, to verify the performance of the RO, we plot an ARC \cite{Sajjad2009}.
The ARC visualizes the trade-off between the classification accuracy and the rejection rate generated by the RO.
Note that the classification accuracy is the top-1 classification accuracy, where the rejected samples are considered correct.
If the AU-ARC is included, it is possible to compare methods regardless of the threshold value \threshold.
The experiments are performed with trained 10-fold cross-validation.

\subsubsection{Machine learning}
In experiments using five different 
trained classification models, linear SVM, multilayer perceptron (MLP), decision tree (DT), random forest (RF), and k-nearest neighbors (k-NN), 
we evaluate the ITRFs and MCFs.
The numbers of samples of the two classes with low numbers are increased to the same level as that of the majority class by SMOTE \cite{Chawla2002}.
The number of neighbors is $5$.
The hyperparameters for each machine learning model are shown below.

The classification model of linear SVM is trained with a regularization parameter $C = 1$.
One-vs-rest (OvR), which can be applied to multiclassification problems, is used.

The MLP in this experiment consists of three layers with a hidden layer of 100 nodes and is trained by backpropagation.
The ramp function is selected as the activation function for all nodes.
Overfitting is suppressed by L2 regularization.
The regularization parameter is set to $ 0.0001 $.
We use Adam \cite{KingmaB14} for the optimization of objective functions.
The Adam parameters are set to the default values in \cite{KingmaB14} ($\alpha =0.0001, \beta_1 =0.9, \beta_2 = 0.99, \epsilon=10^{-8} $).
Training is terminated if the loss value in the training data is not less than 0.0001
 in a minimum of 10 iterations.

The DT is trained by the classification and regression tree (CART) algorithm with the Gini diversity index.
The training is terminated when the number of samples or classes present in all leaves reaches one or when the depth reaches 12.

The RF is composed of 100 decision tree classifiers that are trained in the same way as the DT.

We set $k = 5$ in kNN.

\subsubsection{Prediction score for the RO}
The following shows how the scores are calculated for each machine learning model.
For the OvR linear SVM, the prediction score is the maximum distance from the decision boundary.
MLP uses the maximum value of the prediction probability vector normalized by the softmax function as the prediction score.
DT uses the proportion of the same class as the predicted class among the classes of the training samples in the leaves as the prediction score.
In RF, each decision tree's prediction score is calculated the same way as in DT, and the average value is the prediction score.
For kNN, the score is the proportion of the highest number of classes in the neighborhood.

\subsection{Experimental results}
\subsubsection{Evaluation classification accuracy} \label{subsubsec:ca}
\begin{table*}[t]
  \begin{center}
    \caption{Balanced accuracy (\%) between ITRFs and MCFs.} \label{tb:result-1}
    \begin{tabular}{|l|l|ccccc|} \hline
    Dataset & Features & Linear SVM & MLP & DT & RF & kNN \\ \hline \hline
      \multirow{2}{*}{AAD} & ITRF & $95.15\  (\pm 5.21)$ & $95.88\  (\pm 4.64)$ & $96.0\  (\pm 3.94)$ & \boldmath{ $95.08\  (\pm 4.59)$ } & $96.03\  (\pm 3.52)$ \\
       & MCF & \boldmath{ $97.12\  (\pm 3.15)$ } & \boldmath{ $96.63\  (\pm 2.96)$ } & \boldmath{ $96.87\  (\pm 3.89)$ } & $94.02\  (\pm 3.49)$ & \boldmath{ $96.96\  (\pm 1.66)$ } \\ \hline
      \multirow{2}{*}{MAD} & ITRF & $59.02\  (\pm 7.7)$ & $61.19\  (\pm 8.88)$ & $69.53\  (\pm 8.52)$ & $64.5\  (\pm 7.77)$ & $61.97\  (\pm 8.41)$ \\
       & MCF & \boldmath{ $87.98\  (\pm 5.14)$ } & \boldmath{ $88.12\  (\pm 4.77)$ } & \boldmath{ $83.63\  (\pm 3.63)$ } & \boldmath{ $83.95\  (\pm 6.92)$ } & \boldmath{ $85.84\  (\pm 3.14)$ } \\ \hline
    \end{tabular}
  \end{center}
\end{table*}

Table \ref{tb:result-1} shows the balanced accuracy results of this experiment.
The ``dataset'' column indicates the target dataset.
The ``features'' column shows the features into which the signature of the dataset has been transformed.
The columns with the names of the machine learning methods show the corresponding means and standard deviations of the balanced accuracy.

All machine learning models achieved a balanced accuracy of 95\% or higher in the AAD experiments.
Therefore, we assume that the ITRFs adequately express the if-then rules.
On the other hand, the balanced accuracy of the ITRFs on the MAD is lower than that on the AAD.
Even though the ITRFs can sufficiently follow the if-then rules, they had difficulty in classifying the MAD.

To compare the IRTFs and MCFs, we confirm the experimental results on the MAD.
The results show that the MCFs had a higher balanced accuracy for all machine learning models.
The performance of linear SVM, MLP, DT, RF and kNN improved by 28.96\%, 26.93\%, 14.10\%, 19.45\%, and 23.87\%, respectively.
All the methods showed an improvement of at least 14.10\%.

We evaluate the performance of the MCFs on the AAD.
The results indicate that no machine learning models showed a significant performance improvement.
The WMFs in the MCFs seemed to capture the characteristics of the manual classification well.

\subsubsection{Evaluation of RO performance} \label{subsubsec:ro}
\begin{table*}[t]
  \begin{center}
    \caption{AU-ARC (\%) between ITRFs and MCFs.} \label{tb:result-2}
    \begin{tabular}{|l|l|ccccc|} \hline
    Dataset & Features & Linear-SVM & MLP & DT & RF & kNN \\ \hline \hline
      \multirow{2}{*}{AAD} & ITRF & $99.93\  (\pm 0.07)$ & $99.93\  (\pm 0.06)$ & $99.74\  (\pm 0.2)$ & \boldmath{ $99.9\  (\pm 0.08)$ } & \boldmath{ $99.57\  (\pm 0.21)$ } \\
       & MCF & \boldmath{ $99.95\  (\pm 0.05)$ } & \boldmath{ $99.98\  (\pm 0.02)$ } & \boldmath{ $99.75\  (\pm 0.17)$ } & $99.88\  (\pm 0.09)$ & $99.5\  (\pm 0.27)$ \\ \hline
      \multirow{2}{*}{MAD} & ITRF & $93.26\  (\pm 1.81)$ & $95.18\  (\pm 0.95)$ & $93.41\  (\pm 1.4)$ & $93.8\  (\pm 1.79)$ & $86.52\  (\pm 1.6)$ \\
       & MCF & \boldmath{ $99.45\  (\pm 0.36)$ } & \boldmath{ $99.42\  (\pm 0.36)$ } & \boldmath{ $97.8\  (\pm 0.69)$ } & \boldmath{ $99.36\  (\pm 0.52)$ } & \boldmath{ $92.88\  (\pm 1.66)$ } \\ \hline
    \end{tabular}
  \end{center}
\end{table*}
\begin{figure*}[t]
 \begin{center}
  \begin{tabular}[t]{c}
   \includegraphics[height=3.3cm]{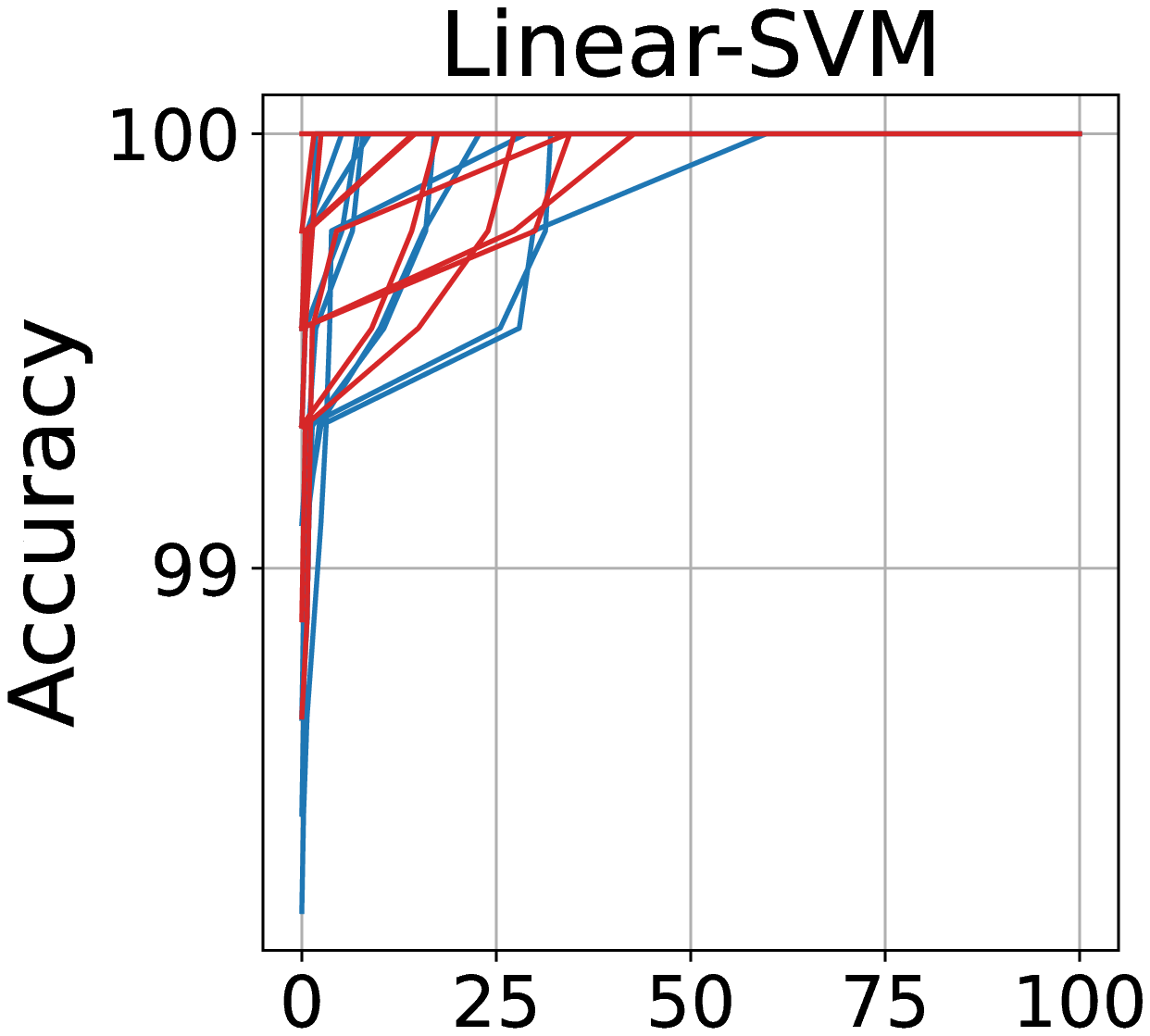}
   \includegraphics[height=3.3cm]{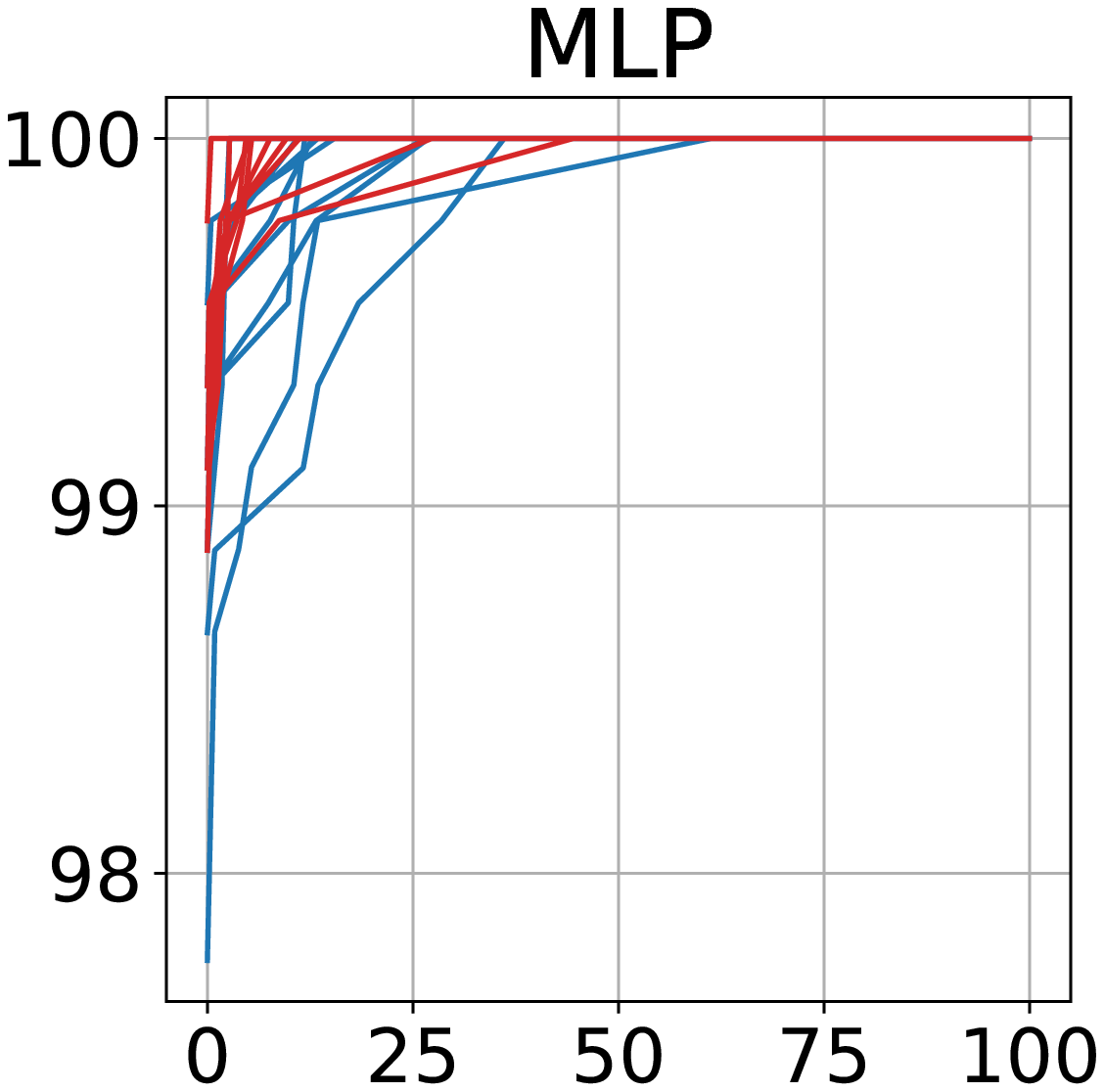}
   \includegraphics[height=3.3cm]{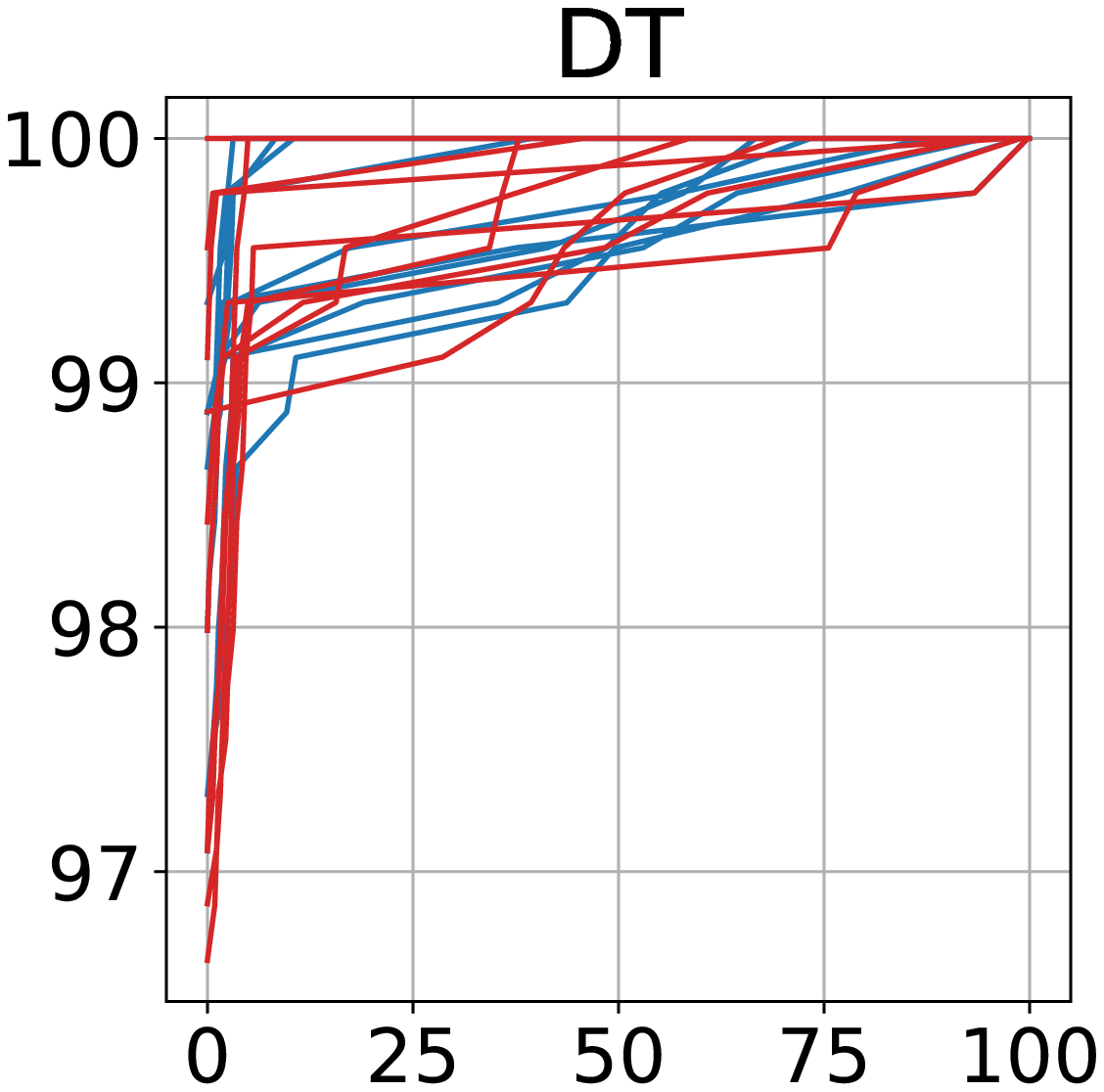}
   \includegraphics[height=3.3cm]{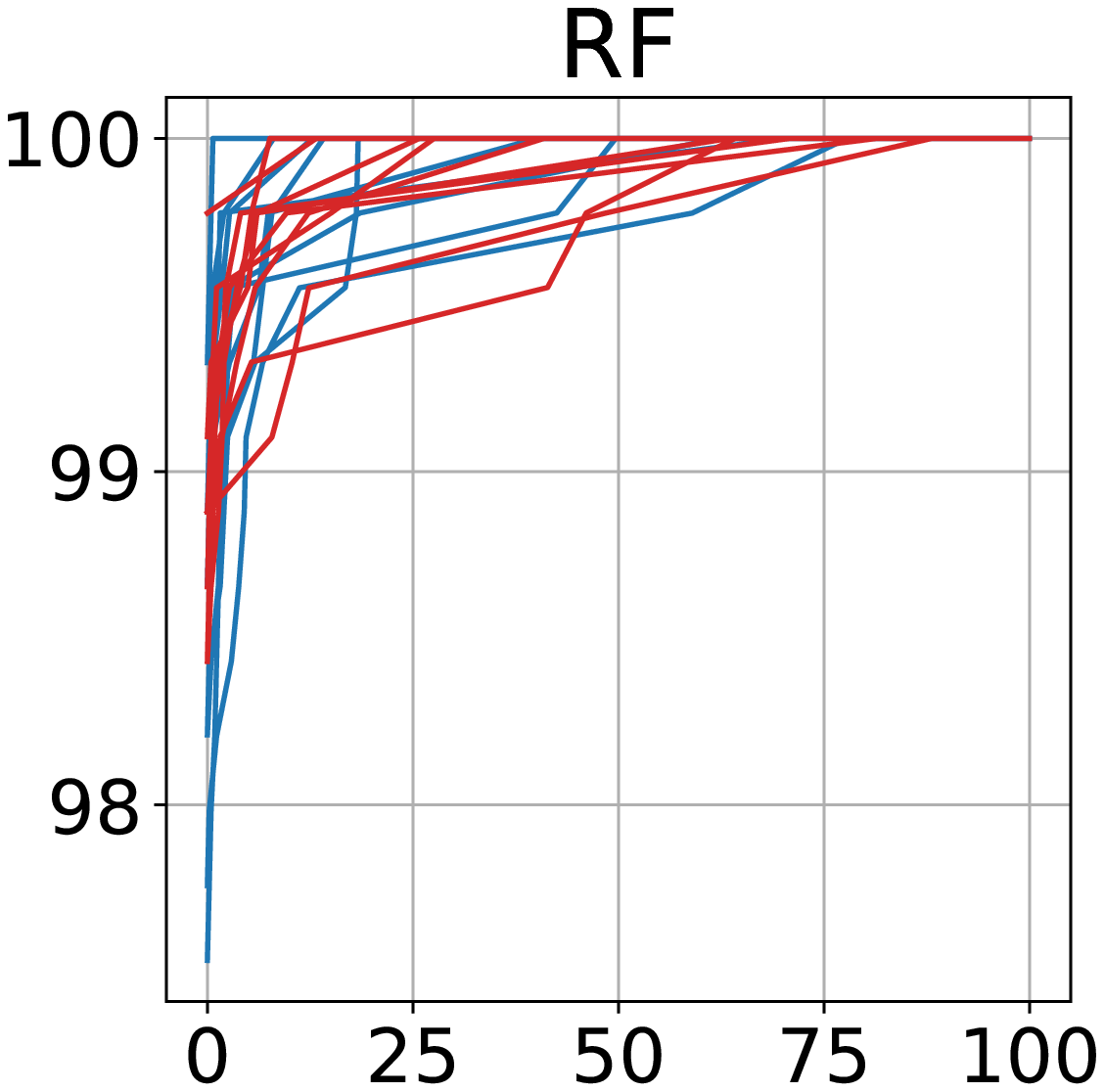}
   \includegraphics[height=3.3cm]{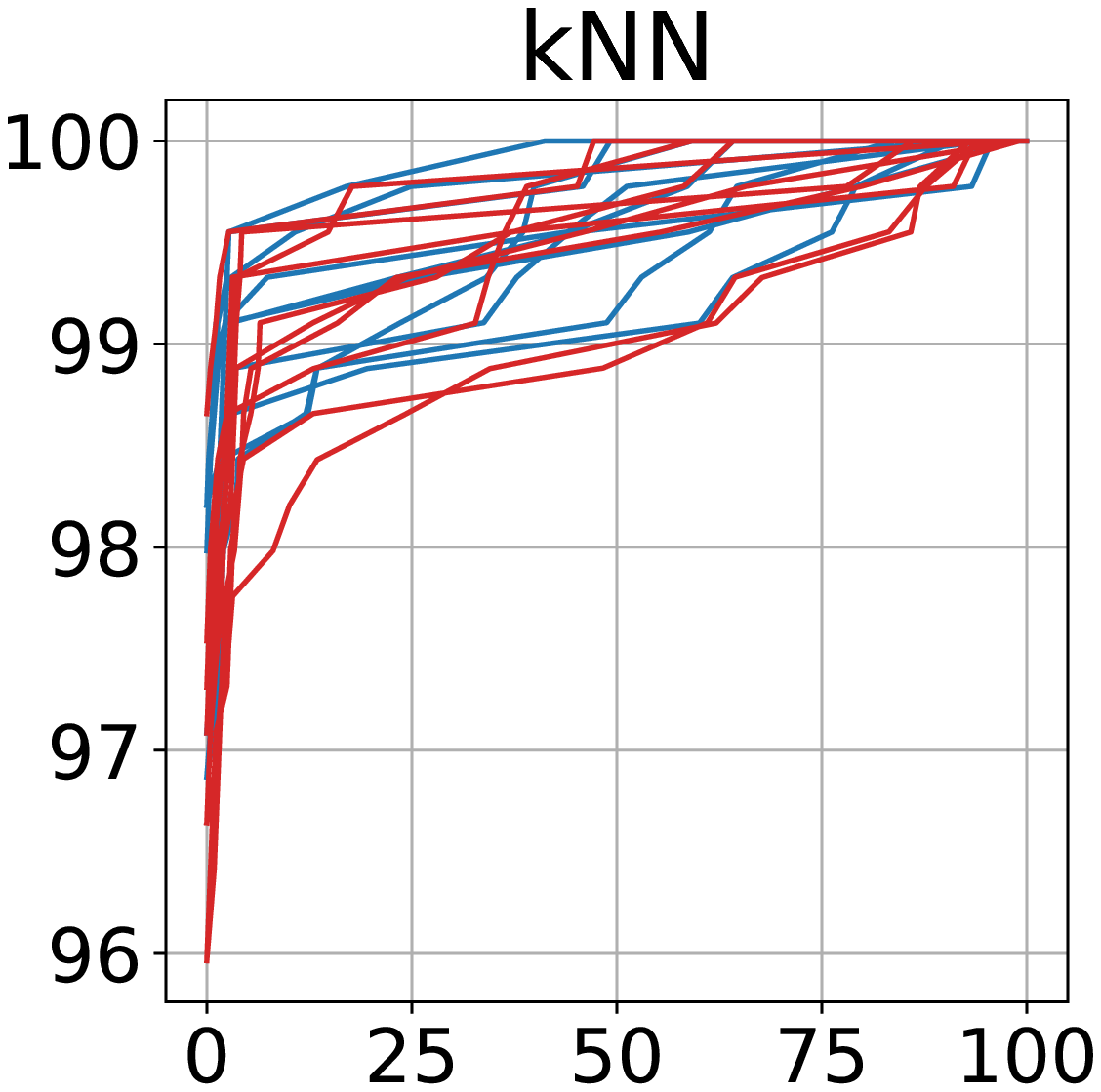}
   \\
   \includegraphics[height=3.3cm]{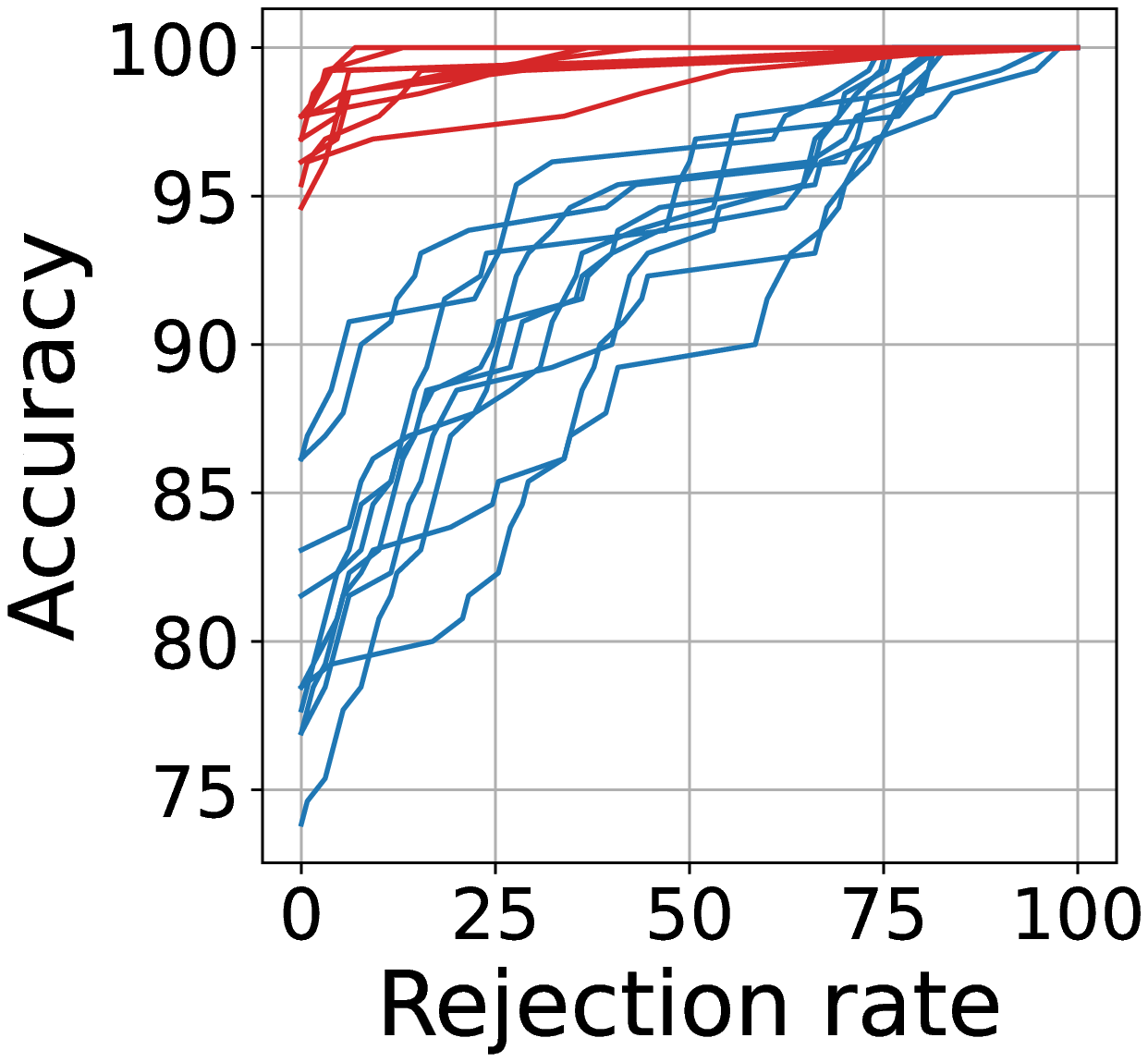}
   \includegraphics[height=3.3cm]{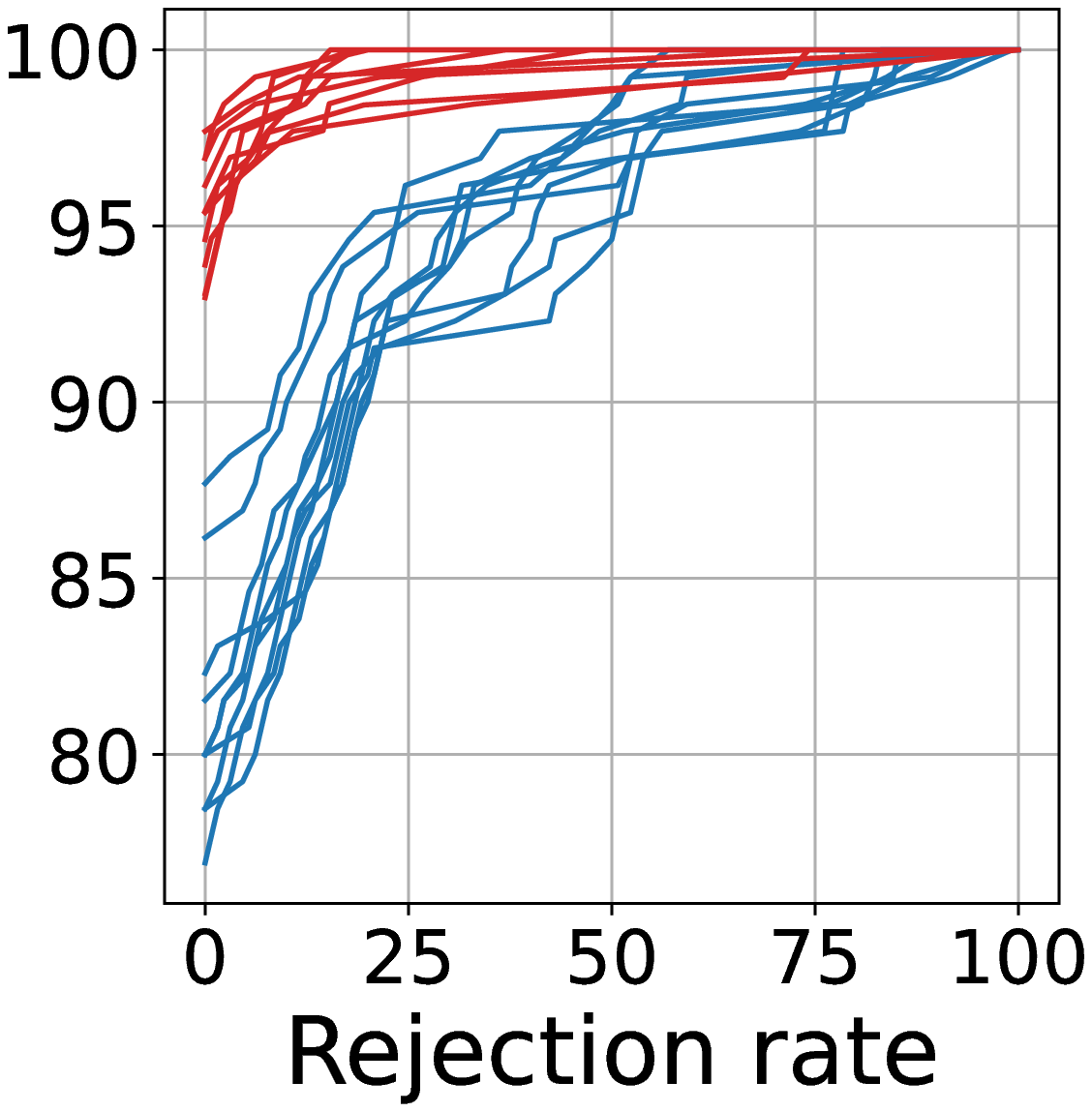}
   \includegraphics[height=3.3cm]{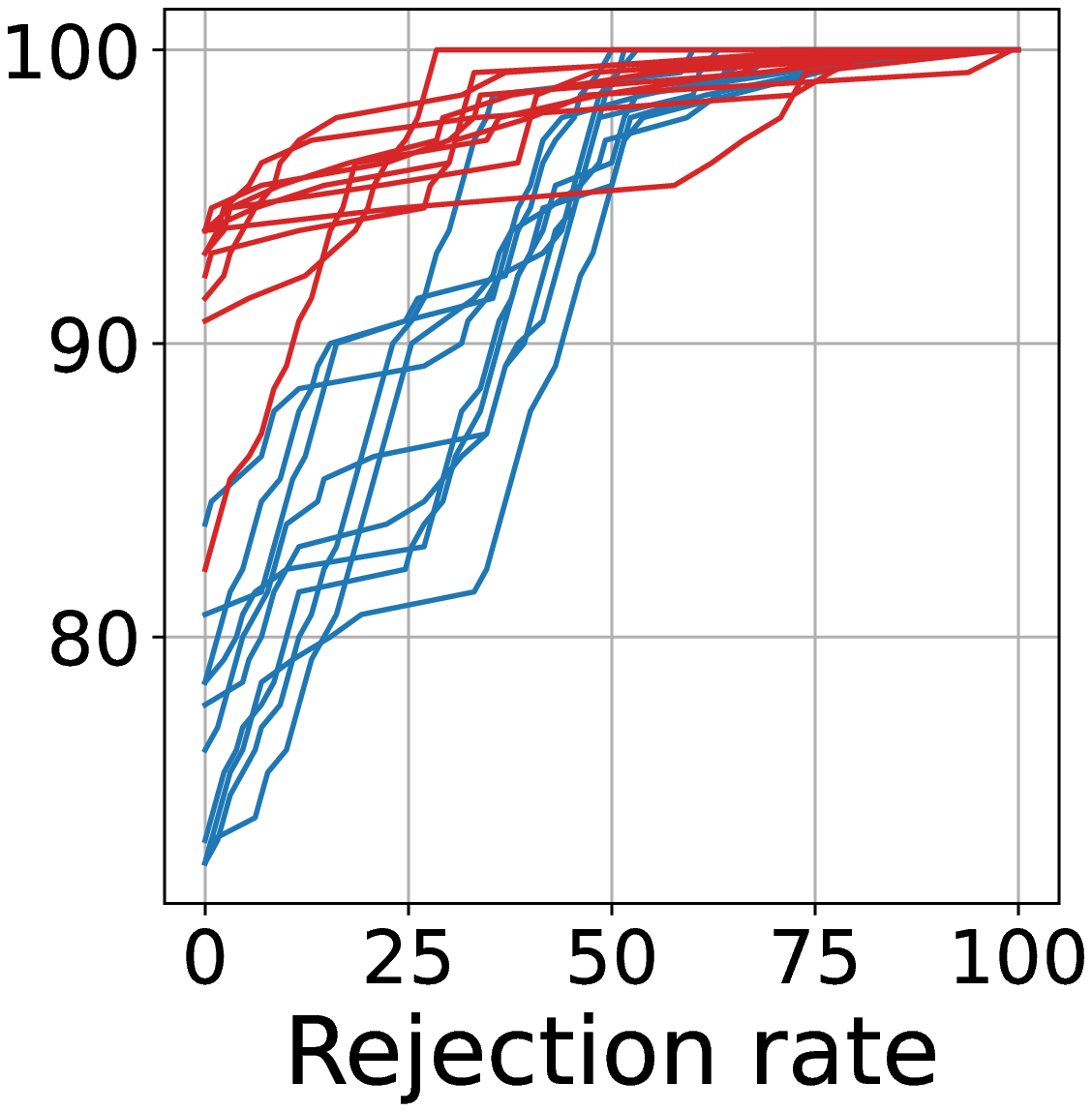}
   \includegraphics[height=3.3cm]{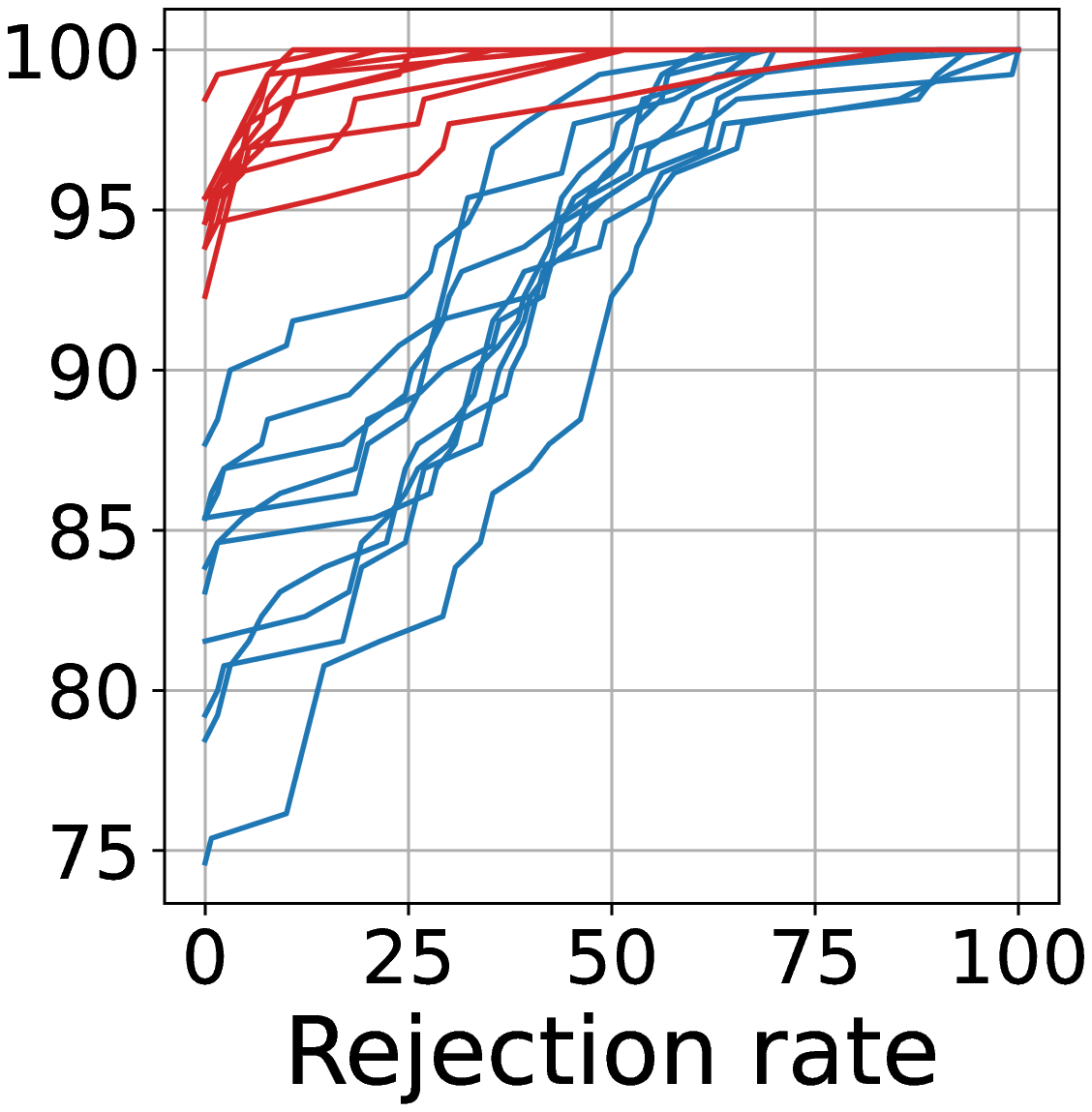}
   \includegraphics[height=3.3cm]{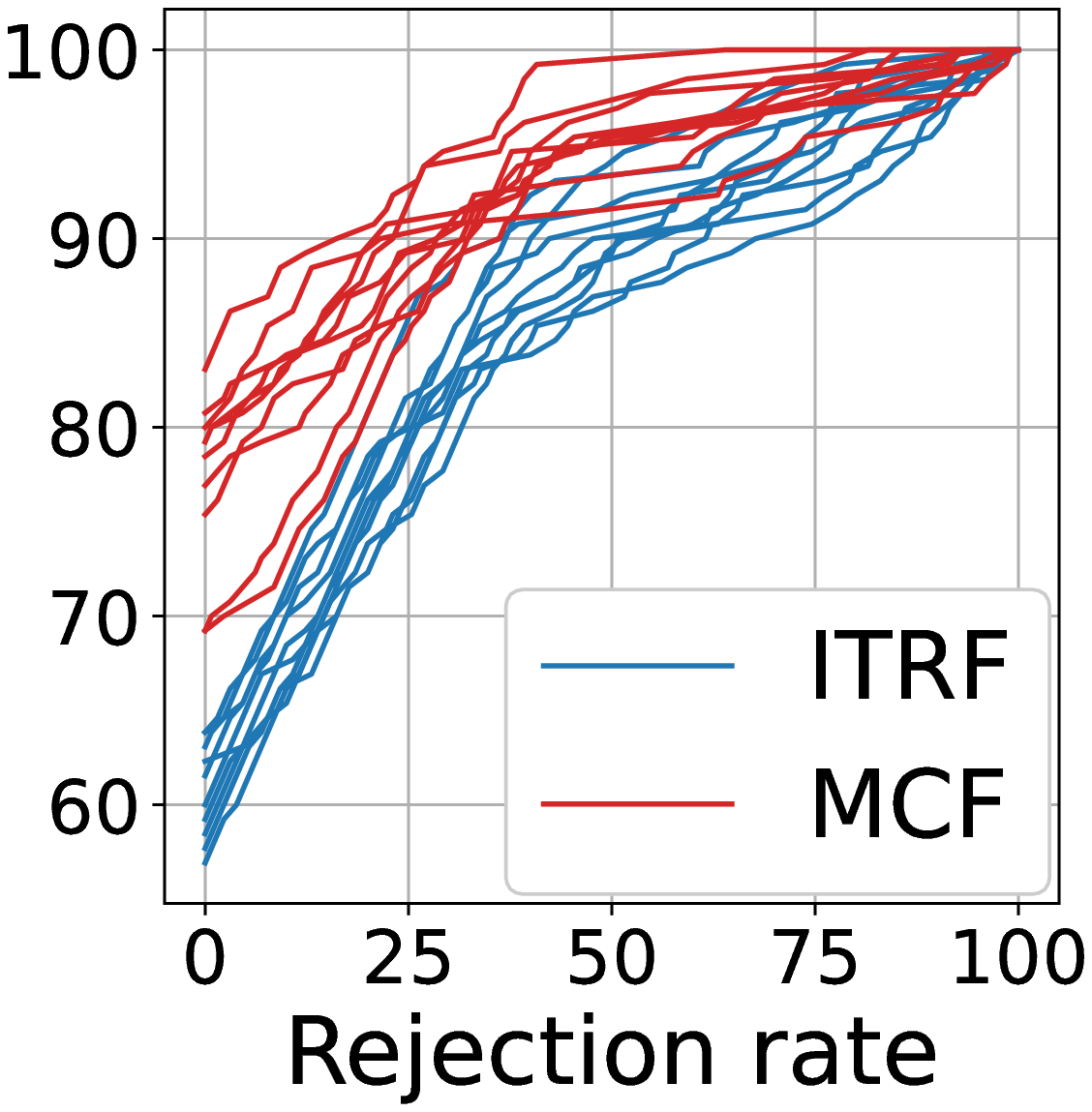}
  \end{tabular}
 \end{center}
 \vspace{-0.2cm}
 \caption{Each ARC shows a 1-fold result of stratified 10-fold cross-validation. The upper panel shows the ARCs on the AAD, and the lower panel shows the ARCs on the MAD. From left to right: linear SVM, MLP, DT, RF, and kNN.} \label{fig:arc}
\end{figure*}

Table \ref{tb:result-2} shows the AU-ARCs.
Fig. \ref{fig:arc} shows all the ARCs for each fold in the stratified 10-fold cross-validation.
The overall trend is similar to that of balanced accuracy.
On the AAD, there was no significant performance difference between the ITRFs and MCFs.
However, on the MAD, the MCFs outperformed the ITRFs for the RO.
The RO performance of linear SVM, MLP, DT, RF and kNN improved by 6.19\%, 4.24\%, 4.39\%, 5.56\%, and 6.36\%, respectively.

In real cases, experts classify signatures that are rejected 
by classification models.
The RO performance on the MAD shows its effectiveness.
On the MAD, the MCFs show a high RO performance of more than 99\% AU-ARC when using linear SVM, MLP, and RF.
This result is one more indication of the practicality of the MCFs.

Comparing the performance between machine learning models on the MAD with MCFs, MLP shows the second-best value, almost equal to that of linear SVM.
To achieve a better trade-off between the rejection rate and accuracy, the accuracy of the estimation of the prediction score is crucial.
Although recent deep neural networks tend to be overconfident \cite{Guo2017}, they are still a better option than other machine learning models in this experiment.
However, the MLP used in this experiment 
performs less well than 
recent DNNs, so overconfidence may not have been a concern.

\subsubsection{Evaluation deep ensemble for the RO} \label{subsubsec:de}
In this section, we use a deep ensemble (DE) \cite{Lakshminarayanan2017} as a classification model to further improve the performance of the RO and confirm its effectiveness.
The RO is said to have a better trade-off between accuracy and rejection rate the closer its prediction score is to the actual probability that it fits the classification \cite{Chow1957}.
DE is considered a method that better represents the uncertainty of deep neural networks (DNNs).
Experiments have also shown the superior calibration capability of DE, which is the ability to estimate the true probability \cite{Lakshminarayanan2017,Ovadia2019}.

\begin{figure}[t]
 \begin{center}
  \begin{tabular}[t]{c}
   \includegraphics[height=3.3cm]{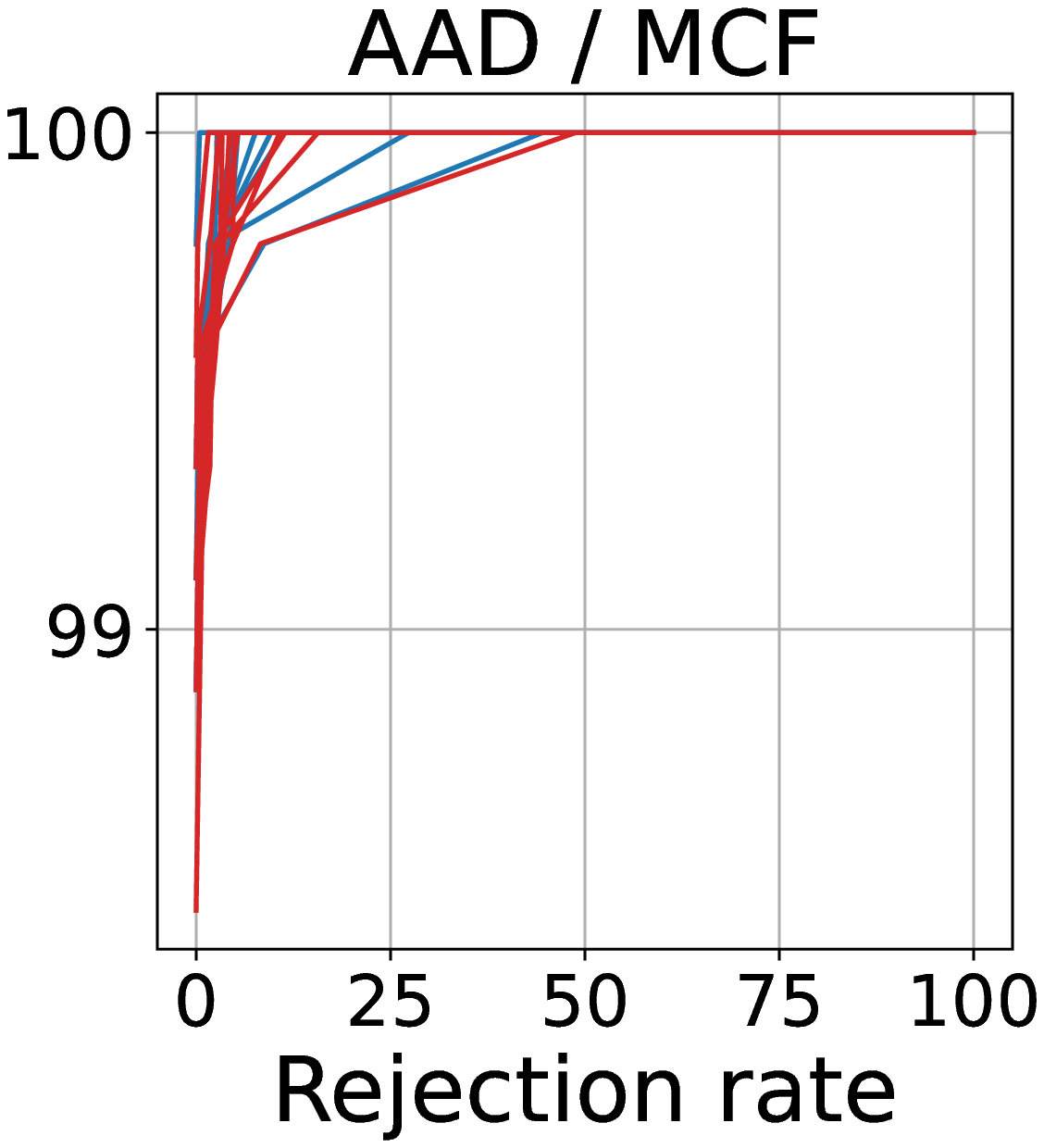}
   \includegraphics[height=3.3cm]{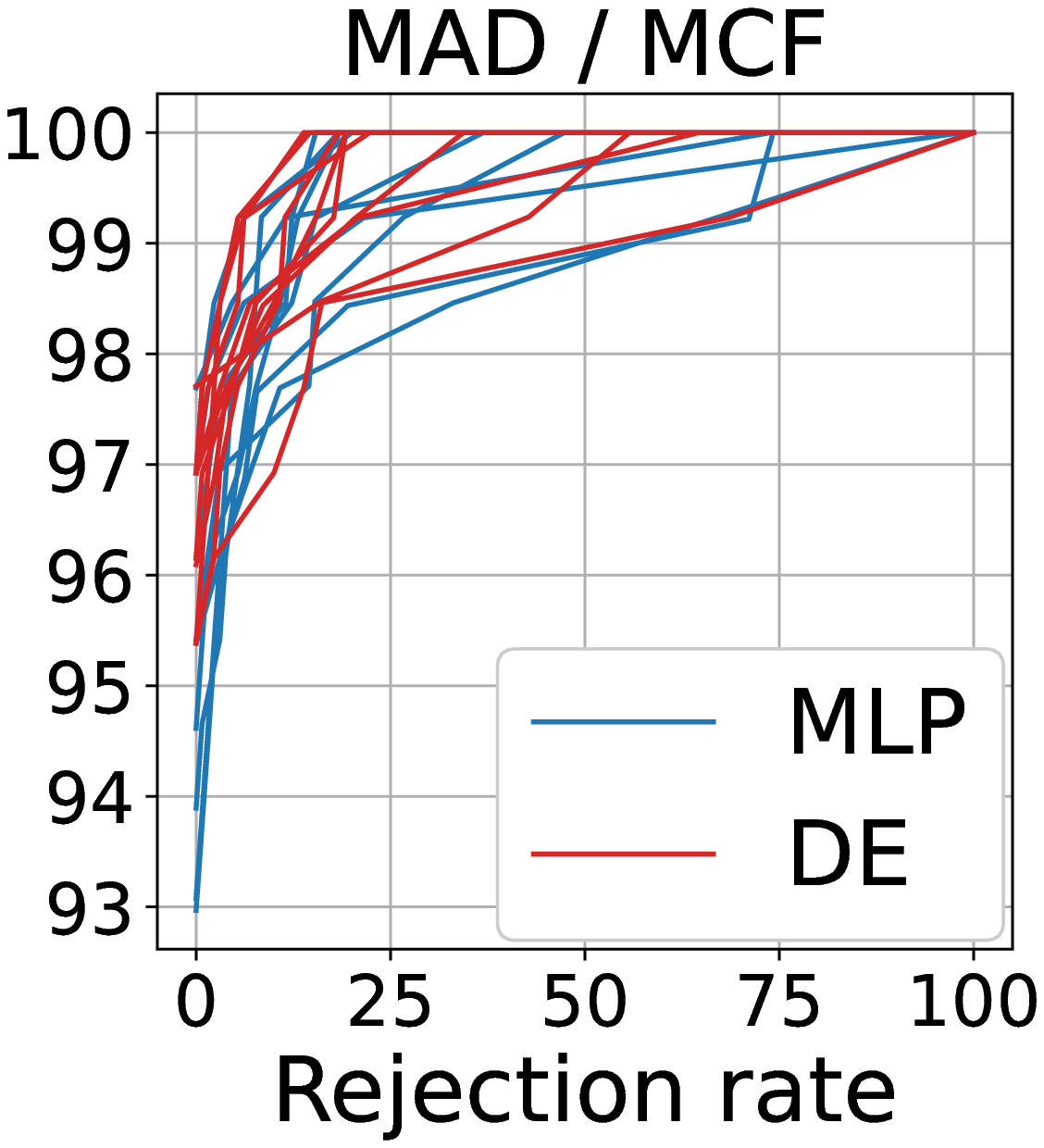}
  \end{tabular}
 \end{center}
 \vspace{-0.0cm}
 \caption{ARC confirms the improvement of the RO with the deep ensemble.}\label{fig:de-arc}
\end{figure}

The analysis results are shown in table \ref{tb:result-3} and Fig. \ref{fig:de-arc}.
Each DE consists of 100 independently trained MLPs, each with identical data.
The prediction score for the RO is the average of the prediction scores output by the component MLPs.
The AU-ARC shows a performance improvement.
The results also show that DE generally 
performs well in terms of the ARC.
The AU-ARC for DE shows the best combined results in table \ref{tb:result-1} and table \ref{tb:result-3}.

DE was also found to positively affect signature classification with the RO.
In our proposed model, any machine learning method can be used for the classification model as long as the RO is feasible.
However, we conclude that DE is the best choice for this experiment.
In addition to the results we measured, the advantage of using DE is that it extends MLP.
MLP is a kind of DNN, and DNNs continue to make remarkable progress in terms of applications.
Therefore, this signature classification model with an RO is also expected to benefit from the future development of MLPs and DNNs.

\begin{table}[t]
  \begin{center}
    \caption{AU-ARC (\%) between MLP and DE.} \label{tb:result-3}
    \begin{tabular}{|l|l|cc|} \hline
    Dataset  & Features & MLP                  & DE \\ \hline \hline
      AAD    & MCF     & $99.98\  (\pm 0.02)$ & $99.98\  (\pm 0.02)$ \\
      MAD    & MCF      & $99.42\ (\pm 0.36)$  & \boldmath{$99.58\  (\pm 0.32)$} \\ \hline
    \end{tabular}
  \end{center}
\end{table}

\subsubsection{Analysis of the expert's point of view} \label{subsubsec:ae}
Through interviews with experts, we designed the WMFs by assuming that all the information in the msg and the information from the web is important.
The values of the weight parameters of the trained models denote the feature importance in classifying signatures, so the features with large weights are considered by experts.
We analyze the feature importance in binary classification tasks (important vs nonimportant), where we merge the high and middle labels of the security importance level into one class (important).
We apply the MCFs and linear SVM to the AAD and MAD for our analysis.
A comparison of the classification model weights learned by linear SVM shows which factors among \textit{5-tuple}, \textit{metadata}, \textit{classtype}, \textit{msg}, and \textit{reference} are considered more important.
The average of the weights learned in each fold of stratified 10-fold cross-validation is calculated.

\begin{figure}[t]
 \begin{center}
    \begin{tabular}[t]{c}
     \hspace{-0.25cm}
   \includegraphics[height=3.3cm]{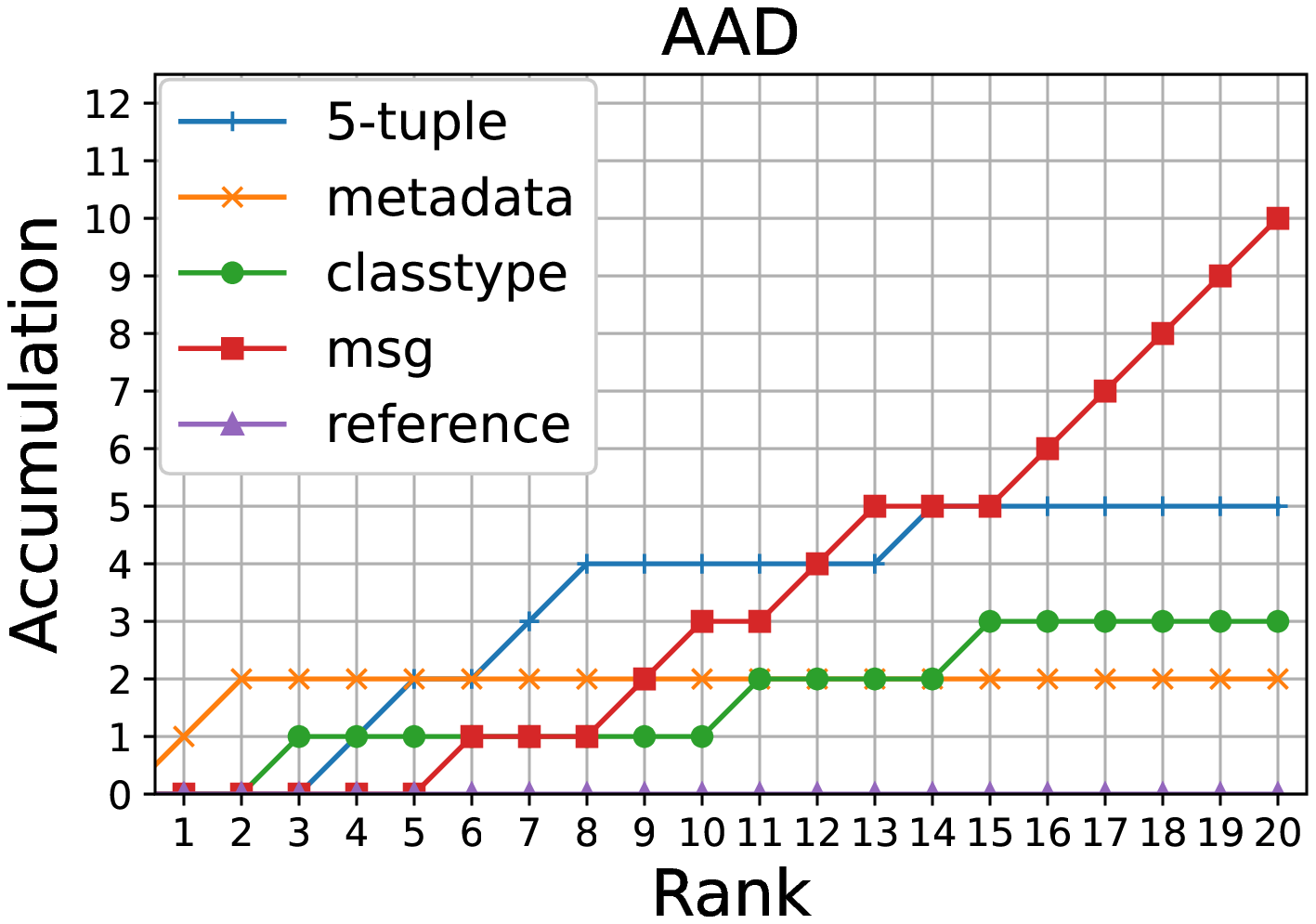}
   \hspace{-0.25cm}
   \includegraphics[height=3.3cm]{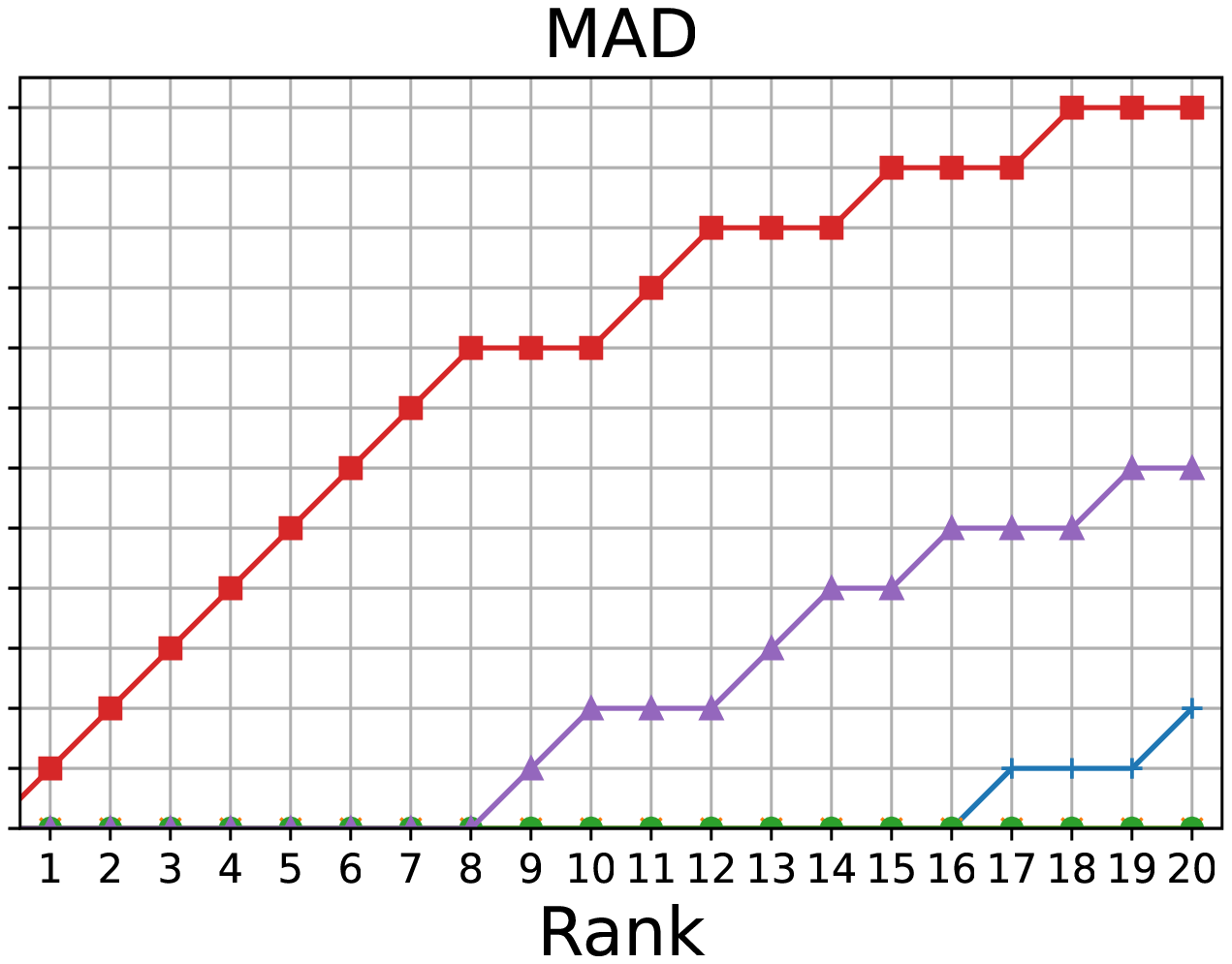}
    \end{tabular}
 \end{center}
 \vspace{-0.0cm}
 \caption{Feature analysis: the cumulative number of elements with high weights for each feature in the ranking on the horizontal axis}\label{fig:importances}
\end{figure}

We can determine to which of the five elements of the signature each weight belongs.
Fig. \ref{fig:importances} shows the cumulative frequency graphs of the classification model weights of the five elements.
The horizontal axis shows the ranks of the absolute values of the weights.
A comparison of the two figures shows that the important features are different depending on the type of dataset.
The best and second-best features in the AAD are elements of \textit{metadata}.
Additionally, in third place is an element of \textit{classtype}, and in fourth place is an element of the 5-tuple.
The elements of \textit{msg} appear in the 6th position and later, but the elements of \textit{reference} do not appear in the top 20.
On the other hand, the weights of the top eight features in the MAD are elements of \textit{msg}.
After the top nine, \textit{reference} elements appear, and after the top 17, 5-tuple elements appear.
\textit{metadata} and \textit{classtype} elements do not appear in the top 20 at all.

The features with high weights were consistent with the features identified as important in the expert interviews.
We find that, unlike if-then rules, experts pay attention to msg and reference in manual classification.
\textit{msg} and the external information from \textit{reference} are similar to natural-language information.
If these are the dominant perspectives in manual classification, then it is likely that natural-language processing (NLP) methods can be applied.
NLP is an area that is advancing rapidly with the rise of deep learning.
For example, we can apply the bidirectional encoder representations from transformers (BERT) model learned from a large corpus of languages \cite{devlin-etal-2019-bert}.
BERT has been applied to many tasks with excellent results \cite{8975793,8959920}.
There is a possibility that it could be applied to improve the classification accuracy of signatures.

\section{Conclusion}
We proposed a machine learning importance classification model with an RO for IDPS signatures.
We designed SFs, KFs, and WMFs as features for this model.
The SFs and KFs were designed based on if-then rules.
The WMFs were designed based on interviews with experts.
We proposed features by combining TF-IDF and web scraping to obtain WMFs.
We developed real datasets, the AAD and MAD, consisting of signatures labeled with importance levels by real experts.
We conducted experiments using five different machine learning models.
The use of combined SFs and KFs classified the AAD with high accuracy.
When we extracted and classified the features from the signatures of the MAD, we obtained only low accuracy.
However, the use of combined SFs and WMFs yielded a significant performance improvement on the MAD.
These trends in the classification accuracy were similar to the performance of the RO.
The results also showed that DE might be promising for improving the performance of the RO.
We also analyzed the validity of the WMFs by checking the weights of the classification model trained by linear SVM.

The experimental results showed that the message text and web-based feature expansion contribute to improving the classification accuracy and RO performance on the MAD.
One future research direction is to use advanced NLP techniques, including statistical language modeling and word embedding.
It is also necessary to explore how to operationalize the classification model to handle the daily generation of signatures.

\bibliographystyle{IEEEtran}

\begin{thebibliography}{10}
\providecommand{\url}[1]{#1}
\csname url@samestyle\endcsname
\providecommand{\newblock}{\relax}
\providecommand{\bibinfo}[2]{#2}
\providecommand{\BIBentrySTDinterwordspacing}{\spaceskip=0pt\relax}
\providecommand{\BIBentryALTinterwordstretchfactor}{4}
\providecommand{\BIBentryALTinterwordspacing}{\spaceskip=\fontdimen2\font plus
\BIBentryALTinterwordstretchfactor\fontdimen3\font minus
  \fontdimen4\font\relax}
\providecommand{\BIBforeignlanguage}[2]{{%
\expandafter\ifx\csname l@#1\endcsname\relax
\typeout{** WARNING: IEEEtran.bst: No hyphenation pattern has been}%
\typeout{** loaded for the language `#1'. Using the pattern for}%
\typeout{** the default language instead.}%
\else
\language=\csname l@#1\endcsname
\fi
#2}}
\providecommand{\BIBdecl}{\relax}
\BIBdecl

\bibitem{Sdas}
S.~{Das} and M.~J. {Nene}, ``A sudrvey on types of machine learning techniques
  in intrusion prevention systems,'' in \emph{International Conference on
  Wireless Communications, Signal Processing and Networking (WiSPNET)}, 2017,
  pp. 2296--2299.

\bibitem{Stakhanova:2010:MID:1752046.1752051}
N.~Stakhanova and A.~A. Ghorbani, ``Managing intrusion detection rule sets,''
  in \emph{Proceedings of Third European Workshop on System Security}, ser.
  EUROSEC '10.\hskip 1em plus 0.5em minus 0.4em\relax ACM, 2010, pp. 29--35.

\bibitem{F5958211}
F.~{Massicotte} and Y.~{Labiche}, ``An analysis of signature overlaps in
  intrusion detection systems,'' in \emph{IEEE/IFIP International Conference on
  Dependable Systems Networks (DSN)}, 2011, pp. 109--120.

\bibitem{H8328450}
H.~{Shahriar} and W.~{Bond}, ``Towards an attack signature generation framework
  for intrusion detection systems,'' in \emph{DASC/PiCom/DataCom/CyberSciTech},
  2017, pp. 597--603.

\bibitem{S7569096}
S.~{Lee}, S.~{Kim}, S.~{Lee}, J.~{Choi}, H.~{Yoon}, D.~{Lee}, and J.~{Lee},
  ``Largen: Automatic signature generation for malwares using latent dirichlet
  allocation,'' \emph{IEEE Transactions on Dependable and Secure Computing},
  vol.~15, no.~5, pp. 771--783, 2018.

\bibitem{N7585840}
N.~{Fallahi}, A.~{Sami}, and M.~{Tajbakhsh}, ``Automated flow-based rule
  generation for network intrusion detection systems,'' in \emph{Iranian
  Conference on Electrical Engineering (ICEE)}, 2016, pp. 1948--1953.

\bibitem{Constantinides}
C.~{Constantinides}, S.~{Shiaeles}, B.~{Ghita}, and N.~{Kolokotronis}, ``A
  novel online incremental learning intrusion prevention system,'' in
  \emph{IFIP International Conference on New Technologies, Mobility and
  Security (NTMS)}, 2019, pp. 1--6.

\bibitem{Chandre}
P.~R. {Chandre}, P.~N. {Mahalle}, and G.~R. {Shinde}, ``Machine learning based
  novel approach for intrusion detection and prevention system: A tool based
  verification,'' in \emph{IEEE Global Conference on Wireless Computing and
  Networking (GCWCN)}, 2018, pp. 135--140.

\bibitem{F8809121}
F.~M. {Cort^^c3^^a9s} and N.~{Gaviria G^^c3^^b3mez}, ``A hybrid alarm
  management strategy in signature-based intrusion detection systems,'' in
  \emph{IEEE Colombian Conference on Communications and Computing (COLCOM)},
  2019, pp. 1--6.

\bibitem{P10.1007/978-3-540-30143-1_6}
T.~Pietraszek, ``Using adaptive alert classification to reduce false positives
  in intrusion detection,'' in \emph{Recent Advances in Intrusion
  Detection}.\hskip 1em plus 0.5em minus 0.4em\relax Springer Berlin
  Heidelberg, 2004, pp. 102--124.

\bibitem{Chow1957}
C.~K. {Chow}, ``An optimum character recognition system using decision
  functions,'' \emph{IRE Transactions on Electronic Computers}, vol. EC-6,
  no.~4, pp. 247--254, 1957.

\bibitem{Chow1970}
C.~{Chow}, ``On optimum recognition error and reject tradeoff,'' \emph{IEEE
  Transactions on Information Theory}, vol.~16, no.~1, pp. 41--46, 1970.

\bibitem{Harish2018}
H.~G. Ramaswamy, A.~Tewari, and S.~Agarwal, ``{Consistent algorithms for
  multiclass classification with an abstain option},'' \emph{Electronic Journal
  of Statistics}, vol.~12, no.~1, pp. 530--554, 2018.

\bibitem{Losing2016}
V.~Losing, B.~Hammer, and H.~Wersing, ``Knn classifier with self adjusting
  memory for heterogeneous concept drift,'' in \emph{IEEE International
  Conference on Data Mining (ICDM)}, 2016, pp. 291--300.

\bibitem{Gomes2017}
H.~M. Gomes, A.~Bifet, J.~Read, J.~P. Barddal, F.~Enembreck, B.~Pfahringer,
  G.~Holmes, and T.~Abdessalem, ``Adaptive random forests for evolving data
  stream classification,'' \emph{Machine Learning}, vol. 106, pp. 1--27, 10
  2017.

\bibitem{Geoepfert2018}
J.~Goepfert, B.~Hammer, and H.~Wersing, ``Mitigating concept drift via
  rejection,'' in \emph{International Conference on Artitificial Neural
  Networks (ICANN)}.\hskip 1em plus 0.5em minus 0.4em\relax Springer, 2018.

\bibitem{Waseem2019}
M.~H. {Waseem}, M.~S.~A. {Nadeem}, A.~{Abbas}, A.~{Shaheen}, W.~{Aziz},
  A.~{Anjum}, U.~{Manzoor}, M.~A. {Balubaid}, and S.~{Shim}, ``On the feature
  selection methods and reject option classifiers for robust cancer
  prediction,'' \emph{IEEE Access}, vol.~7, pp. 141\,072--141\,082, 2019.

\bibitem{Lin2018}
D.~Lin, L.~Sun, K.~Toh, J.~Zhang, and Z.~Lin, ``Biomedical image classification
  based on a cascade of an svm with a reject option and subspace analysis,''
  \emph{Computers in Biology and Medicine}, vol.~96, pp. 128--140, 2018.

\bibitem{Raghu2019}
M.~Raghu, K.~Blumer, R.~Sayres, Z.~Obermeyer, B.~Kleinberg, S.~Mullainathan,
  and J.~Kleinberg, ``Direct uncertainty prediction for medical second
  opinions,'' in \emph{International Conference on Machine Learning}, ser.
  Proceedings of Machine Learning Research, K.~Chaudhuri and R.~Salakhutdinov,
  Eds., vol.~97.\hskip 1em plus 0.5em minus 0.4em\relax PMLR, 2019, pp.
  5281--5290.

\bibitem{A8723825}
A.~I. {Kadhim}, ``Term weighting for feature extraction on twitter: A
  comparison between bm25 and tf-idf,'' in \emph{International Conference on
  Advanced Science and Engineering (ICOASE)}, 2019, pp. 124--128.

\bibitem{Y8253040}
Y.~{Yang}, ``Research and realization of internet public opinion analysis based
  on improved tf - idf algorithm,'' in \emph{International Symposium on
  Distributed Computing and Applications to Business, Engineering and Science
  (DCABES)}, 2017, pp. 80--83.

\bibitem{P8250358}
P.~{Sun}, L.~{Wang}, and Q.~{Xia}, ``The keyword extraction of chinese medical
  web page based on wf-tf-idf algorithm,'' in \emph{International Conference on
  Cyber-Enabled Distributed Computing and Knowledge Discovery (CyberC)}, 2017,
  pp. 193--198.

\bibitem{nothman-etal-2018-stop}
J.~Nothman, H.~Qin, and R.~Yurchak, ``Stop word lists in free open-source
  software packages,'' in \emph{Workshop for {NLP} Open Source Software
  ({NLP}-{OSS})}, 2018, pp. 7--12.

\bibitem{Sajjad2009}
M.~S.~A. Nadeem, J.-D. Zucker, and B.~Hanczar, ``Accuracy-rejection curves
  (arcs) for comparing classification methods with a reject option,'' in
  \emph{International Workshop on Machine Learning in Systems Biology}, ser.
  Proceedings of Machine Learning Research, vol.~8, 2009, pp. 65--81.

\bibitem{Lakshminarayanan2017}
B.~Lakshminarayanan, A.~Pritzel, and C.~Blundell, ``Simple and scalable
  predictive uncertainty estimation using deep ensembles,'' in \emph{Advances
  in Neural Information Processing Systems (NeurIPS)}, 2017, pp. 6402--6413.

\bibitem{Chawla2002}
N.~V. Chawla, K.~W. Bowyer, L.~O. Hall, and W.~P. Kegelmeyer, ``Smote:
  Synthetic minority over-sampling technique,'' vol.~16, no.~1, pp. 321--357,
  Jun. 2002.

\bibitem{KingmaB14}
D.~P. Kingma and J.~Ba, ``Adam: {A} method for stochastic optimization,'' in
  \emph{International Conference on Learning Representations (ICLR)}, Y.~Bengio
  and Y.~LeCun, Eds., 2015, pp. 1--15.

\bibitem{Guo2017}
C.~Guo, G.~Pleiss, Y.~Sun, and K.~Q. Weinberger, ``On calibration of modern
  neural networks,'' in \emph{International Conference on Machine Learning
  (ICML)}, 2017, pp. 1321--1330.

\bibitem{Ovadia2019}
Y.~Ovadia, E.~Fertig, J.~Ren, Z.~Nado, D.~Sculley, S.~Nowozin, J.~Dillon,
  B.~Lakshminarayanan, and J.~Snoek, ``Can you trust your model's uncertainty?
  evaluating predictive uncertainty under dataset shift,'' in \emph{Advances in
  Neural Information Processing Systems (NeurIPS)}, 2019, pp. 13\,991--14\,002.

\bibitem{devlin-etal-2019-bert}
J.~Devlin, M.-W. Chang, K.~Lee, and K.~Toutanova, ``{BERT}: Pre-training of
  deep bidirectional transformers for language understanding,'' in \emph{Annual
  Conference of the North {A}merican Chapter of the Association for
  Computational Linguistics}, 2019, pp. 4171--4186.

\bibitem{8975793}
W.~{Li}, S.~{Gao}, H.~{Zhou}, Z.~{Huang}, K.~{Zhang}, and W.~{Li}, ``The
  automatic text classification method based on bert and feature union,'' in
  \emph{IEEE International Conference on Parallel and Distributed Systems
  (ICPADS)}, 2019, pp. 774--777.

\bibitem{8959920}
Y.~{Iwasaki}, A.~{Yamashita}, Y.~{Konno}, and K.~{Matsubayashi}, ``Japanese
  abstractive text summarization using bert,'' in \emph{International
  Conference on Technologies and Applications of Arti^^ef^^ac^^81cial
  Intelligence (TAAI)}, 2019, pp. 1--5.

\end{thebibliography}

\end{document}